\documentclass[aps,prd,final,twocolumn,letterpaper, superscriptaddress]{revtex4}

\usepackage{amsmath, amssymb, amsthm, cancel, dsfont, enumerate, epstopdf, eucal, fancyhdr, feynmf, gensymb, graphicx, lipsum, mathtools, physics, tikz, verbatim}
\usepackage[caption=false]{subfig}

\usepackage{hyperref}
\hypersetup{
    pdfnewwindow=true,     
    colorlinks=true,     
    linkcolor=blue,          
    citecolor=blue,      
    filecolor=blue,   
    urlcolor=blue    
}

\def\VEV#1{\left\langle #1 \right\rangle}

\begin{document}

\title{Seeking Neutrino Emission from AGN through Temporal and Spatial Cross Correlation}
\author{Cyril Creque-Sarbinowski}
\email{creque@jhu.edu}
\author{Marc Kamionkowski}
\email{mkamion1@jhu.edu}
\author{Bei Zhou}
\email{beizhou@jhu.edu}
\affiliation{Department of Physics and Astronomy, Johns Hopkins University, 3400 N. Charles St., Baltimore, MD 21218, USA}
\date{\today} 

\begin{abstract}
Active galactic nuclei (AGN) are a promising source for
high-energy astrophysical neutrinos (HEANs). By the end of
2022, the Vera C.\ Rubin
Observatory (VRO) will begin to observe $\gtrsim10$ million AGN
with a regular and high cadence. Here, we evaluate the capacity
of VRO, in tandem with various current and upcoming neutrino
telescopes, to establish AGN as HEAN emitters.  To do so,
we assume that the neutrino luminosity from any given AGN at any
given time is proportional to the electromagnetic luminosity.
We then estimate the
error with which  this fraction can be measured through spatial
and temporal cross-correlation of VRO light curves with IceCube,
KM3NeT, and Bakail-GVD.  We find that it may be possible to
detect AGN contributions at the $\sim3 \sigma$ level to the HEAN flux even if these AGN
contribute only $\sim10\%$ of the HEAN flux. The bulk of this
information comes from spatial correlations, although the
temporal information improves the sensitivity a bit.  The results also imply that if an angular correlation is detected with high signal-to-noise, there may be prospects to detect a correlation between AGN variability and neutrino arrival times.
The small HEAN fraction estimated here to be accessible to the
entirety of the VRO AGN sample suggests that valuable
information on the character of the emitting AGN may be obtained
through similar analyses on different sub-populations of AGN.
\end{abstract}

\maketitle

\pagestyle{myheadings}
\markboth{Cyril Creque-Sarbinowski}{Seeking Neutrino Emission from AGN through Temporal and Spatial Cross Correlation}
\thispagestyle{empty}

\section{Introduction}
High-energy astrophysical neutrinos (HEANs) comprise a diffuse
isotropic extragalactic background of neutrinos observed with
energies between a few TeV to a few PeV~\cite{1908.09551,
2001.09520, 2011.03545, 2101.09836}. There is some evidence of
an association of some these neutrinos with the blazar TXS
0506+056~\cite{1807.08794, 1807.08816}, but the source of the
vast majority of the HEAN background remains a mystery. Various
classes of bright AGN population have been constrained to
contribute no more than a fraction of the total observed HEAN
flux~\cite{1502.03104,  1702.08779, 1807.04748, 1904.06371,
2007.12706, 2103.12813}, but there is little known about the
possible contribution of the many lower-luminosity AGN. With the
advent of the Vera Rubin Observatory at the end of 2022, at
least 10 million AGN will be observed in the southern sky with
high cadence for the following 10 years~\cite{0912.0201}. In
addition, neutrino telescopes KM3NeT and
Baikal-GVD will soon be completed in the northern hemisphere
with comparable volume and better angular resolution than
IceCube~\cite{1601.07459, 1808.10353}. Due to their locations,
these telescopes will detect HEAN from upgoing tracks
originating from the southern sky without contamination from the
atmospheric neutrino background. Thus, over the next decade of
AGN and neutrino observations, we expect a large increase in
sensitivity in the determination of AGN as HEAN emitters. 

AGN are hypothesized to emit high-energy neutrinos through either
hadronuclear~\cite{1306.3417, 1807.05113} or
photohadronic~\cite{astro-ph/9802280, 1307.2793}
processes. Therefore, one avenue of examination is the modelling
of these processes under various AGN environments. High-energy
neutrinos can be produced from radio-loud AGN
jets~\cite{Mannheim:1995mm, 1403.4089, 1711.03757}, tidal
disruption events~\cite{2005.08937}, blazar inner cores and
jets~\cite{astro-ph/9501064, 1807.04748}, and AGN coronae
~\cite{1904.04226,1909.02239, 2102.11921, 2105.08948}.  They may
also have nothing to do with AGN---e.g., they may be associated
with choked supernova jets~\cite{1512.08513, 1706.02175, 1809.09610, 2107.09317} and even
cosmic strings~\cite{1108.2509,1206.2924}. Even without
theoretical modeling, information about the source of neutrinos
may be sought with the coincidence of neutrino events with
various source events through data alone~\cite{1405.7648,
1406.0376, 1408.3664, 1501.05158, 1507.05711, 1509.02444,
2001.00930, 2002.06234, 2007.12706, 2103.12813}.
 
In this work we present a statistical framework to determine
whether HEANs are produced by AGN and assess its potential in
the context of VRO, IceCube, KM3NeT, and Baikal-GVD. More
specifically, we propose to cross-correlate temporal and spatial
data from AGN variability and neutrino events.  To evaluate the
prospects to detect such a cross-correlation, we make the
simplest assumption that the neutrino flux from any given AGN at
any given time is proportional to the electromagnetic flux at
that given time.  We then use state-of-the-art information on
the AGN redshift/luminosity distribution and variability
parameters to forecast the detectability of this
cross-correlation.  We find, with the AGN population assumed,
that a correlation can be established even if the AGN in VRO
contribute as little as a few percent to the HEAN flux.  Most of
the sensitivity comes from angular information; the temporal
information contributes approximately 10\% of the signal to
noise.  Our estimates suggest that the better angular resolution
($\sim0.2^\circ$) expected for Baikal-GVD and KM3NeT, relative
to the $\sim0.5^\circ$ for IceCube, will give them roughly twice
the sensitivity to an AGN-neutrino cross-correlation for equal
exposure. 

This paper is organized as follows. In Sec.~\ref{sec:form} we
present the formalism of the AGN/neutrino
cross-correlation.   We discuss our model for the AGN
redshift/luminosity distribution and the variability properties
of AGN in Sec.~\ref{sec:agn_pop}.  We provide and discuss
numerical results in Sec.~\ref{sec:4cast}.  We discuss these
results and conclude in Secs.~\ref{sec:disc} and~\ref{sec:conc}, respectively. 

\section{Formalism}\label{sec:form}

Our aim will be to determine the fraction $f$ of neutrinos that
come from AGN in the sample under the hypothesis that the
neutrino flux from any given AGN at any given time is
proportional to the electromagnetic flux from that AGN at that
given time.  To begin, we will simplify by neglecting AGN
variability and then generalize later.

\subsection{Angular information only}

The optimal estimator to determine the fraction $f$ of neutrinos
that come from AGN in the sample will be the unbinned
maximum-likelihood estimator
\cite{1012.2137,1307.6669,0801.1604,0912.1572,2103.12813},
\begin{equation}
     {\cal L}(f;{\rm data}) = \Pi_i \left[ f S_i + (1-f) B_i \right],
\end{equation}
where the product is over all neutrino events.  Here, $S_i=S(\vec \theta_i) d^2\vec\theta$ is the probability that a given source neutrino will be found in a differential area $d^2\theta$ centered at the position $\theta_i$ of the $i$th neutrino, and $B_i=B(\vec \theta_i) d^2\theta_i$ the analogous quantity for a background neutrino.  We normalize both distributions such that their integral over the area $4\pi f_{\rm sky}$ of the survey (where $f_{\rm sky}$ is the fraction of the sky surveyed).  Then, $B_i=(4\pi f_{\rm sky})^{-1}$, and we take the signal probability to be
\begin{equation}
     S_i = \sum_\alpha w_\alpha
     \frac{1}{2\pi \sigma^2} \exp \left( -\frac{
     (\vec\theta_i -\vec\theta_\alpha )^2}{2 \sigma^2}
     \right).
\end{equation}     
The sum on $\alpha$ is over all AGN in the sample; $\vec\theta_\alpha$ is the position of the $\alpha$th AGN; and $\sigma$ is the error in the neutrino
angular position.  Here, $w_\alpha$ is the probability that a given signal neutrino comes from the $\alpha$th AGN, and so $\sum_\alpha w_\alpha=1$.

Even if there is no signal, the likelihood will most generally be maximized at a value of $f$ selected from a distribution with a variance $\sigma_f^2$ determined from 
\begin{equation}
    \frac{1}{\sigma_f^2} = \left\langle \left(\frac{\partial \ln {\cal L}(f;{\rm data})}{\partial f} \right)^2 \right\rangle,
\end{equation}
where the derivative is evaluated at $f=0$, and the average is taken over all realizations of the data under the null hypothesis.  We use 
\begin{equation}
    \left( \frac {\partial \ln{\cal L}}{\partial f}\right)_{f=0} = \sum_i \left(\frac{S_i}{B}-1 \right).
\end{equation}
We then evaluate the expectation value
\begin{equation} 
    \left\langle \left( \frac{S}{B}-1 \right)^2 \right\rangle = \frac{4 \, f_{\rm sky}}{\sigma^2\,N_{\rm AGN}} \frac{\langle w_\alpha^2 \rangle}{\langle w_\alpha \rangle^2}
\end{equation}
from the zero-lag correlation function for a collection of $N_{\rm AGN}$ randomly distributed AGN.  Here, the angle brackets denote an average over all AGN in the
sample.  Our hypothesis here is that the neutrino flux from AGN $\alpha$ is proportional to its electromagnetic $F_\alpha$, and so the correlation of neutrinos with AGN can be established with a signal-to-noise,
\begin{equation}
     \frac{S}{N} \simeq 2 \frac{f}{\sigma}
     \sqrt{\frac{N_\nu}{N_{\rm AGN}} \frac {\VEV{
     F_\alpha^2}}{\VEV{F_\alpha}^2} f_{\rm sky}}.
\label{eqn:centralresult}     
\end{equation}     
The result has the expected
scalings with $f_{\rm sky}$, $\sigma$, $N_\nu$, and $N_{\rm
AGN}$.  The signal to noise is also weighted by the ratio 
$\VEV {F_\alpha^2}/\VEV{ F_\alpha}^2$ of the
second moment of the AGN flux distribution to the square of the
first moment (i.e., average flux).

\subsection{Including AGN variability}

\begin{figure}[h!!]
\subfloat{\includegraphics[width = \linewidth]{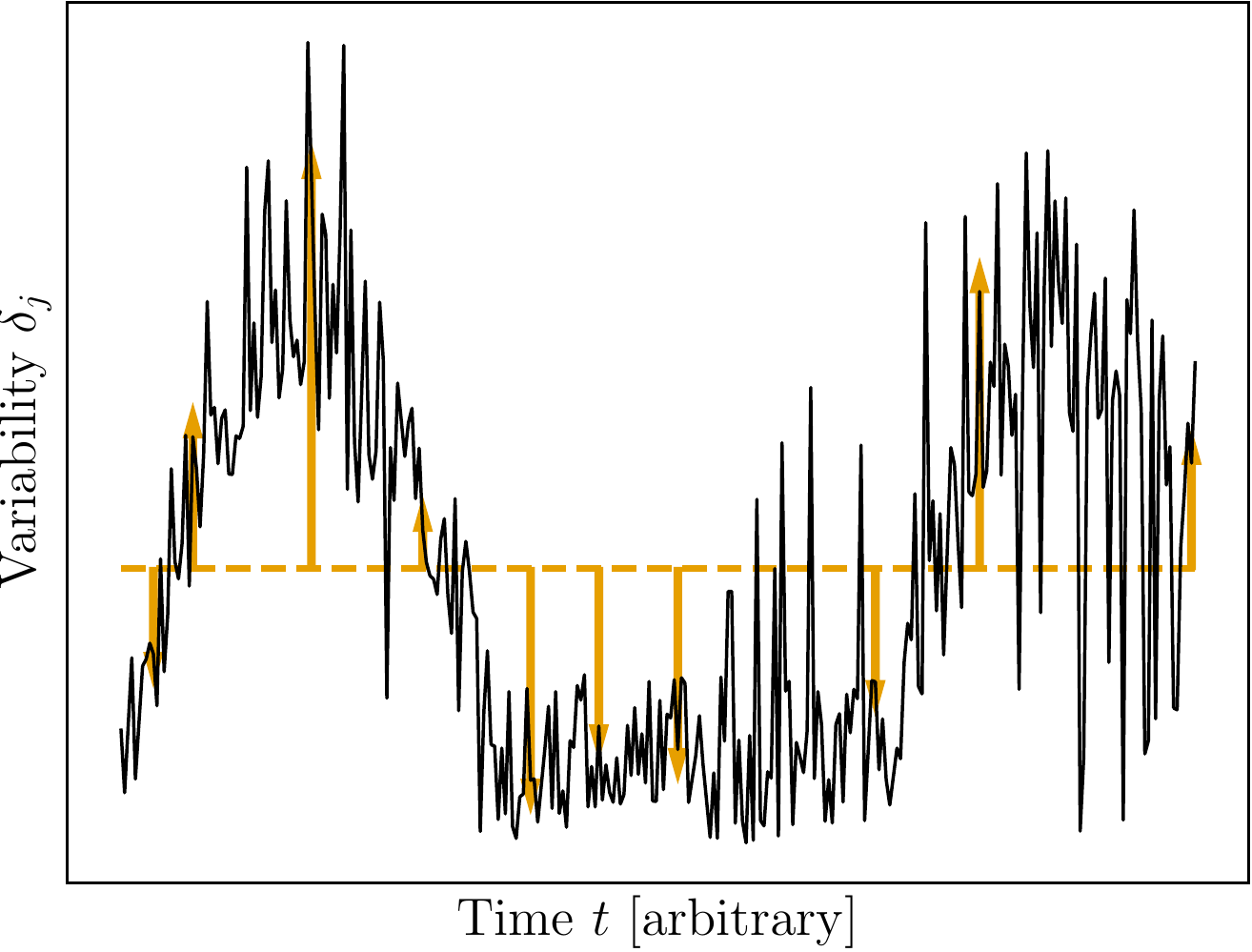}}
\qquad
\subfloat{\includegraphics[width = \linewidth]{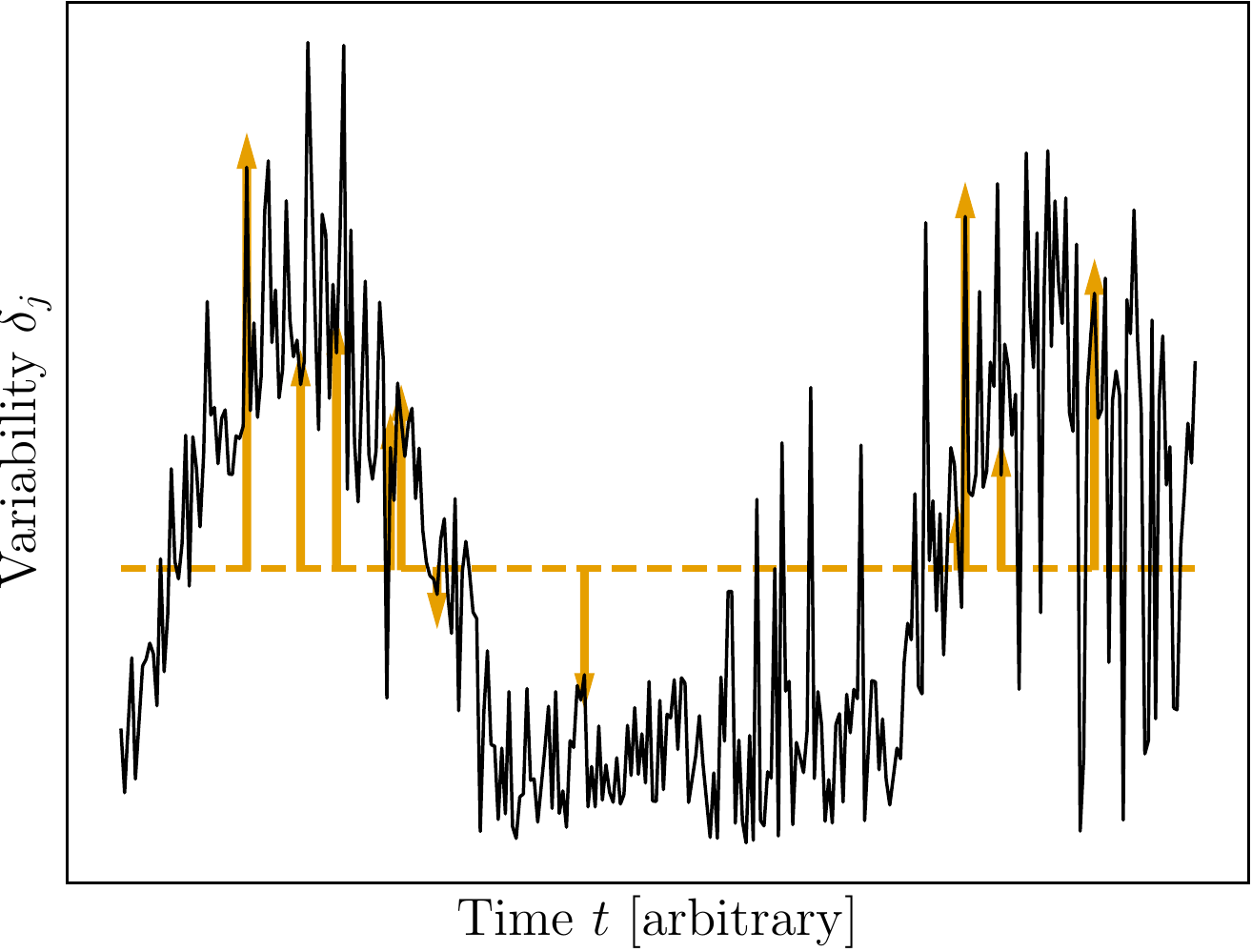}}
\caption{An example of the variability $\delta_j(t)$ of AGN $j$
shown in solid black, plotted against the variability
$\delta_j(t_\alpha)$ evaluated at the neutrino arrival times
$t_\alpha$, depicted by solid orange arrows. The dashed orange
line indicates the zero point, $\delta_j = 0$. In the top figure
the neutrinos arrive randomly, and thus a cross correlation
between these two quantities would become zero. In the bottom
figure the neutrinos are sourced from AGN and thus are biased
towards appearing when the intensity is higher, leading to
nonzero correlation.}\label{fig:temporal}
\end{figure}

The derivation above can be easily generalized to include the
additional information provided by cross-correlating the
neutrino arrival times with AGN luminosity at any given time.
If the neutrino luminosity at any given time is correlated
with the AGN luminosity at that same time, then the probability
to detect a neutrino when a given AGN is, say, twice as bright,
should be twice as large, as illustrated in
Fig.~\ref{fig:temporal}. 

To incorporate this correlation into the likelihood analysis
above, we simply assume that the flux $F_\alpha$ for any given
AGN in the likelihood function is the apparent flux {\it at the
neutrino arrival time}.  The signal-to-noise contributed by
each AGN is then enhanced by a factor,
\begin{equation}
     \left[\frac{\VEV{F^2(t)}}{\VEV{F(t)}^2}\right]^{1/2},
\end{equation}
where here $\VEV{F(t)}$ is the time-averaged flux and
$\VEV{F^2(t)}$ the time-averaged squared flux.  Thus, for
example, if an AGN has a sinusoidal flux variation, $F(t) = F_0
+ F_1 \cos\omega t$, the contribution of this AGN to the total
signal-to-noise is enhanced by a factor $\left[1+(F_1/F_0)/2
\right]^{1/2}$.

Of course, AGN variability is complicated and poorly
understood.  Below we describe a model for AGN variability but
here note that the signal-to-noise with which $f$ can be
inferred will be enhanced by a factor $[1+ \VEV{\sigma_{\rm
var}^2}/2]^{1/2}$, where $\VEV{\sigma_{\rm var}^2}$ is an
appropriately weight rms fractional flux variation.

\section{The AGN Population}\label{sec:agn_pop}

\subsection{The flux distribution}

Here we model the flux distribution we expect for AGN in VRO.
Let $dN_{\rm AGN}/dz dL$ be the redshift and bolometric
luminosity distribution of AGN. Then AGN are distributed
throughout the Universe according to 
\begin{align}\label{eq:dFdLdz}
\frac{dN_{\rm AGN}(z, L)}{dz dL} &= \frac{dV(z)}{dz}\frac{dn(z, L)}{dL},
\end{align}
with $dn(z, L)/dL$ the AGN luminosity function, $dV(z)/dz= 4\pi
f_{\rm sky} r(z)^2 dr(z)/dz$ the comoving volume observed over a
fraction $f_{\rm sky}$ of the sky, $r(z) = \int_0^z |dr/dz| dz$
the comoving radial distance to a redshift $z$, $dr/dz = - c/(1
+ z)H(z)$ its redshift derivative, $c$ the speed of light, and
$H^2(z) = H_0^2\left[\Omega_m(1 + z)^3 + (1 -
\Omega_m)\right]^{-1/2}$ the Hubble parameter. We use Planck
2018 $\Lambda$CDM parameters $H_0 = 2.18\times 10^{-18} {\rm
s}^{-1}$ and $\Omega_m = 0.315$~\cite{1807.06209}, along with
the Full AGN luminosity function in Table 3 from
Ref.~\cite{astro-ph/0605678}. 

However, given a cosmological distribution of AGN, only those
that appear bright enough will be observed. More specifically,
given a limiting apparent magnitude $m_{\rm lim}$, the
distribution of observed AGN in that band is
\begin{align}\label{eq:dFdmdz}
\frac{dN_{\rm AGN}}{dz dm} &= \Theta(m_{\rm lim} - m)\frac{dL_{\rm bol}}{dm}\frac{dN_{\rm AGN}[z, L_{\rm bol}(z, m_b)]}{dz dL_{\rm bol}},
\end{align}
with $\Theta(x)$ the Heaviside theta function, $L_{\rm bol}(z,
m) = K_o(m)4\pi d_L(z)^2\langle\delta\nu_b\rangle F_{\rm
AB}10^{-(2/5)m_b}$ the bolometric luminosity for an AGN with
apparent magnitude $m$ averaged over frequency bands $b$ located
at redshift $z$, and $dL_{\rm bol}/dm = -(2/5)\log(10)L_{\rm
bol}(m, z)$ its apparent magnitude derivative. Furthermore,
$K_o(m)$ is the bolometric correction function to convert from
the emitted luminosity in the optical band to the bolometric
luminosity of the source, $d_L(z) = (1 + z)r(z)$ the luminosity
distance, $\delta\nu_b$ the frequency bandwidth of band $b$, and
$F_{\rm AB} = 3.631\times 10^{-23}\ {\rm W}\ {\rm Hz}^{-1}\ {\rm
m}^{-2}$. Moreover, we assume that the observed intensity is
roughly constant across the entire frequency bandwidth, and that
any redshifting effects on the frequency do not alter the
intensity in each band significantly. While the bolometric
correction is typically a function of the apparent magnitude, it
only varies up to $20\%$ within the optical band across the
magnitudes considered. Thus, for simplicity, we adopt that
$K_o(m) = 10$ for all magnitudes and
bands~\cite{astro-ph/0605678}.

\begin{figure}
\includegraphics[width = 0.48\textwidth]{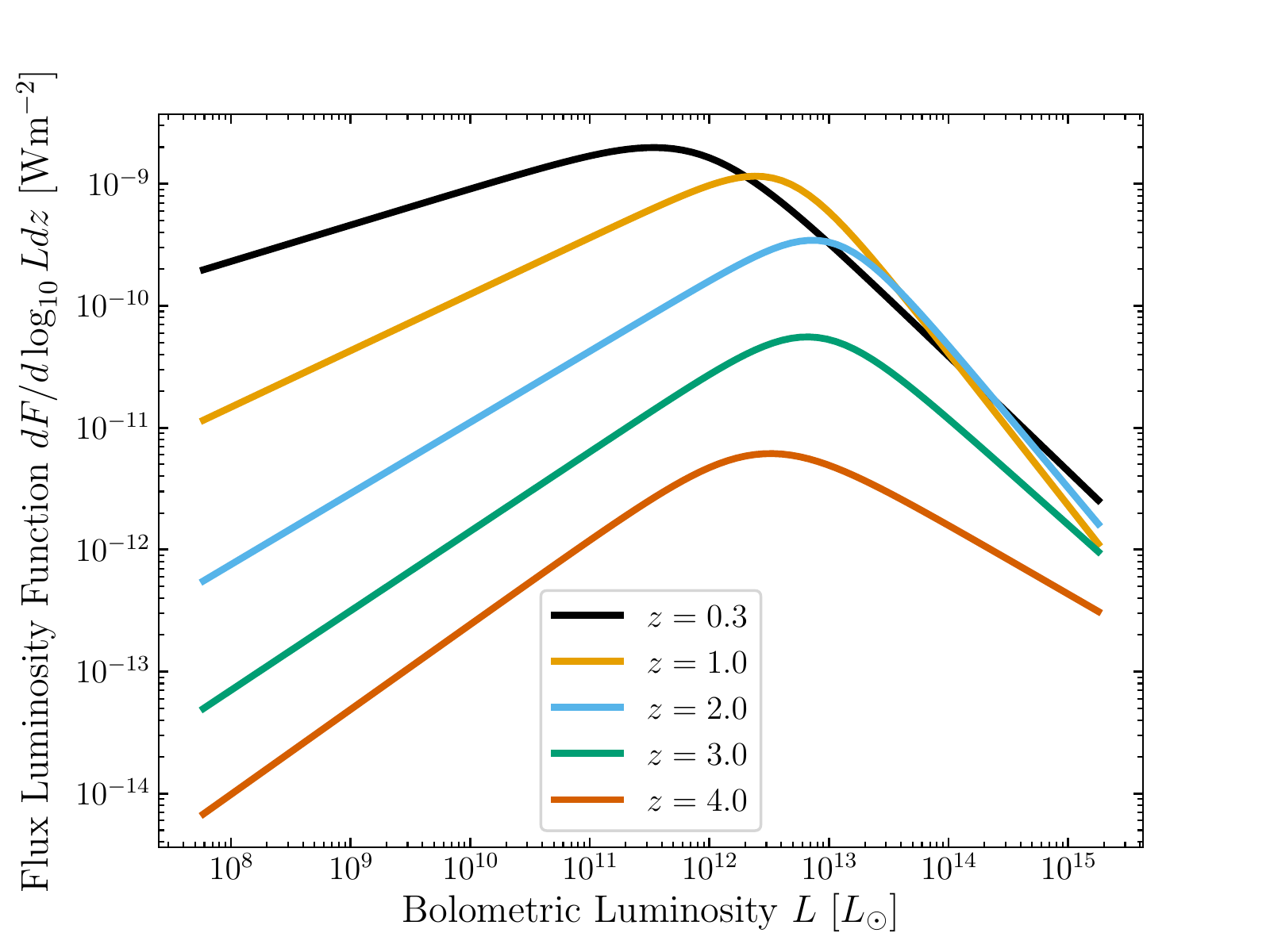}
\caption{The flux distribution of AGN as a function of an AGN's bolometric luminosity at redshifts $z \in \{0.3, 1.0, 2.0, 3.0, 4.0\}$, as given by Eq.~\eqref{eq:dFdLdz}.}\label{fig:dFdLdz}.
\end{figure}

\begin{figure}
\includegraphics[width = 0.48\textwidth]{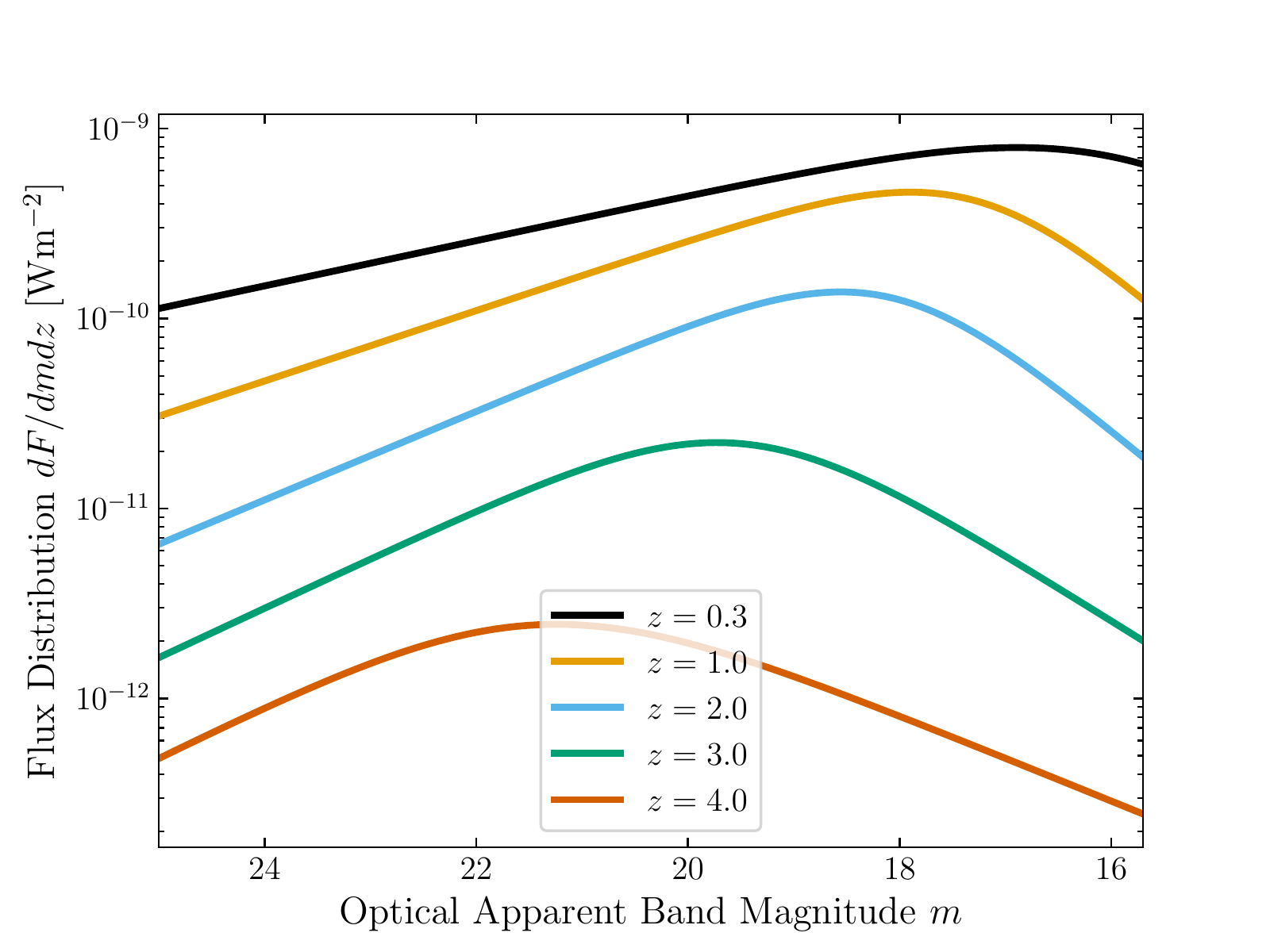}
\caption{The flux distribution of AGN as a function of an AGN's apparent magnitude in an optical band $b$ at redshifts $z \in\{0.3, 1.0, 2.0, 3.0, 4.0\}$, as given by Eq.~\eqref{eq:dFdmdz}. In order to show the full range of this distribution, we do not include the theta function factor.}\label{fig:dFdmdz}
\end{figure}

We define $dF_{\rm AGN}/dz dL$ to be
the AGN flux luminosity distribution, and $dF_{\rm AGN}^b/dzdm
\equiv F dN_{\rm AGN}^b/dz dm$ its magnitude counterpart.
We plot the flux luminosity distribution in
Fig.~\ref{fig:dFdLdz} and the magnitude distribution in
Fig.~\ref{fig:dFdmdz}.

\begin{figure}
\includegraphics[width = 0.48\textwidth]{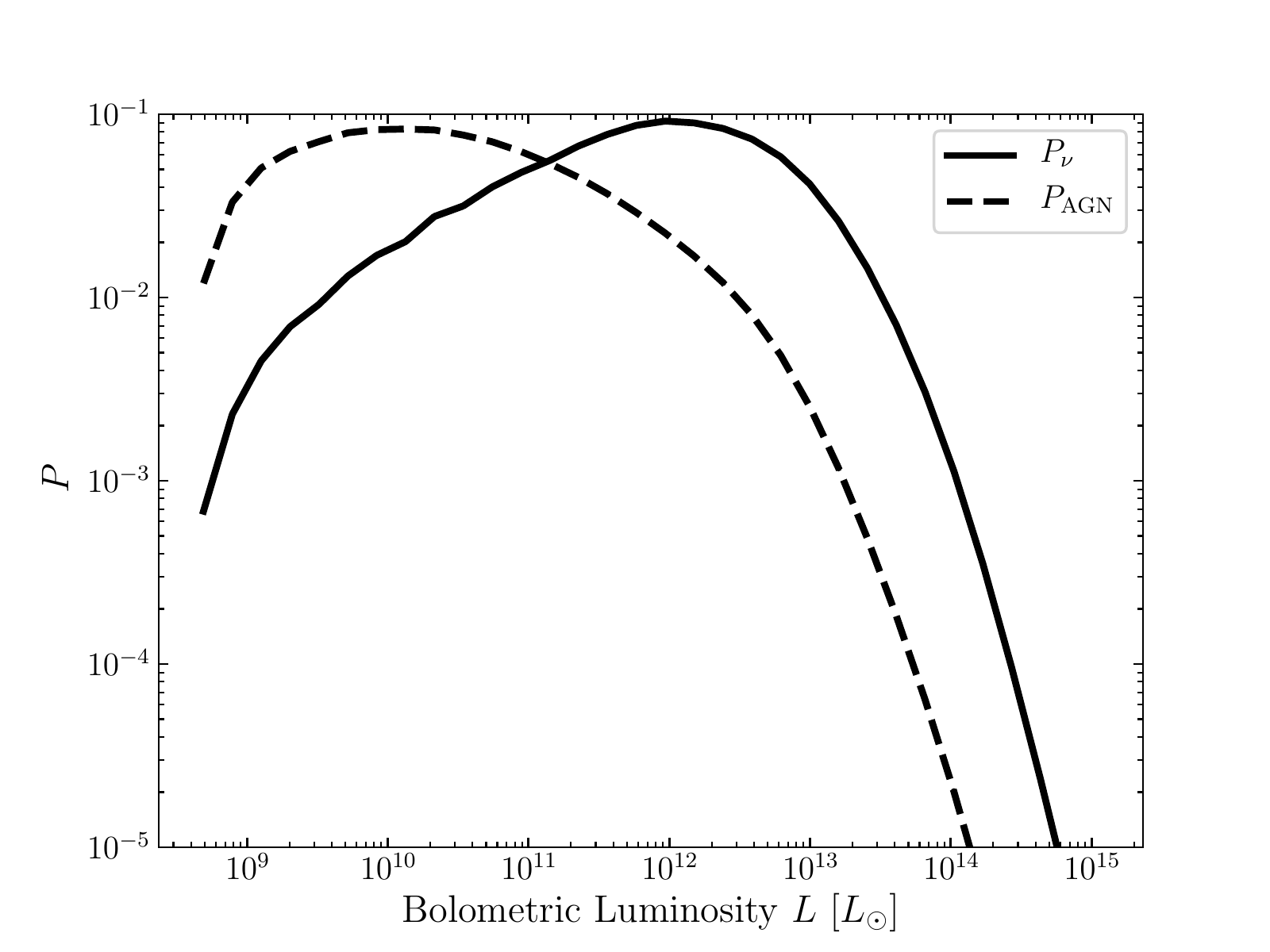}
\caption{The observation probability $P$ associated with
neutrinos and AGN in each bolometric luminosity bin between
redshifts $0.3 \leq z \leq 6.7$. Over the range of luminosities presented,
33 base-10 logarithmic bins are taken. The solid black line is
the probability that an AGN emitting neutrinos has bolometric
luminosity $L$, while the dashed black line is the probability
of that an observed AGN has bolometric luminosity $L$. Both
curves are normalized by the set of observed AGN and are
computed assuming a limiting magnitude of $m_{\rm lim} =
24.0$. At large luminosities both curves follow the expected
flux distribution curve of Fig.~\ref{fig:dFdLdz}, however at
small $L$ the limiting magnitude restricts the total number of
AGN observed.}\label{fig:PvL}
\end{figure}

In Fig~\ref{fig:PvL} we plot the bolometric-luminosity
distribution of AGN in the forecast VRO sample and also the
bolometric-luminosity weighted by the luminosity---this latter
quantity is then proportional to the probability, under our
assumptions, that a given neutrino comes from an AGN of some
given luminosity.

\subsection{AGN Variability}
\label{sec:agnvariability}

The study of AGN variability is in its infancy when compared to how it will appear in the VRO era \cite{ccs2021}.  We thus
have far less in the way of precise current knowledge to
make forecasts for the possibility to detect the cross-correlation
between AGN variability and neutrino arrival times.  To do so,
though, we assume AGN light curves undergo a
damped random walk \cite{1004.0276, 1112.0679, 1202.3783,
1604.01773, 1611.08248, 1701.00005, 1909.06374, 1607.04299, 2108.05389},
a model that provides a reasonable
description of most light curves.  In this case, the intensity
$I_\alpha(t)$ of AGN $\alpha$ undergoes fluctuations described by a
stationary random process with two-point correlation function,
$\VEV{ I_\alpha(t+t')I_\alpha(t)} = A_\alpha^2 e^{-t/\bar
t_\alpha}$, or equivalently, a power spectrum $P_\alpha(\omega)
= 2 A_\alpha^2 \bar t_\alpha/[1+(\omega \bar t_\alpha)^2]$.  We assume
that AGN all have the same
variability amplitude $A_\alpha=1$ and an observer-frame
variability timescale $\bar t_\alpha = \bar t_0(1+z)
(L/L_b)^\beta$ for an AGN of luminosity $L$ at redshift $z$, as
suggested by recent measurements \cite{2108.05389}.  We take $\bar t_0= 1\ {\rm month}$,
$L_b=2\times10^{35}\ {\rm W}$, and $\beta=0.23$.   Our calculation then discards 
Fourier modes with periods longer or shorter than those accessed
by VRO.

Given this population of AGN, we can write the fractional flux variation as 
\begin{equation}
\sigma_{\rm var}^2 = \frac{1}{\pi}\frac{1}{\left\langle F_\alpha^2\right\rangle}\int_{\omega_{\rm min}}^{\omega_{\rm max}} d\omega \left\langle F_\alpha^2 P_\alpha(\omega)\right\rangle,
\end{equation}
where only modes between $\omega_{\rm min} = 2\pi/T$ and $\omega_{\rm max} = 2\pi/\Delta t$ are included, with $\Delta t = 3.5\ {\rm days}$ the temporal resolution of the experiment and $T = 10\ {\rm years}$ the duration of the observation.

\section{Forecasts}\label{sec:4cast}

We now forecast the ability of the neutrino telescopes IceCube,
KM3NeT, and Baikal-GVD, along with optical telescope VRO, to
determine the fraction $f$ of neutrinos that come from AGN in
the survey.

\subsection{Angular information only}

With our model for the AGN luminosity/redshift distribution and
VRO's apparent-magnitude cutoff, we forecast $N_{\rm AGN} \simeq
2.8\times 10^7$ AGN in the survey and 
$\VEV{F_\alpha^2}/\VEV{F_\alpha}^2 \simeq 15$.  We then find
from Eq.~(\ref{eqn:centralresult}),
\begin{eqnarray}
     \frac{S}{N} &\simeq & 5.7\, f \left( \frac{
     (N_\nu/10^4)}{(\sigma/0.5^\circ)(N_{\rm AGN}/2.8\times
     10^7)}\right)^{1/2} \nonumber \\
     & & \times \left( \frac{
     \VEV{F_\alpha^2}/\VEV{F_\alpha}^2}{15} \right)^{1/2} \left(\frac{f_{\rm
     sky}}{0.5} \right)^{1/2}.
\label{eqn:centralresultwithnumbers}     
\end{eqnarray}

\subsection{Angular information and timing}

With our models for AGN variability and the AGN
luminosity/redshift distribution, we infer an rms fractional
flux variation of $\VEV{\sigma_{\rm var}^2}\simeq 0.54$.  The
estimate in Eq.~(\ref{eqn:centralresultwithnumbers} is thus
enhanced by approximately 12\%.

This calculation can also be understood in a different way.  It
suggests that if AGN are determined from angular information to
contribute a fraction $f$ of the observed neutrinos, then a
correspondence between instantaneous AGN luminosity and neutrino
luminosity can be established with a signal-to-noise of
$\left[\VEV{\sigma_{\rm var}^2}/2 \right]^{1/2}$ times the value
in Eq.~(\ref{eqn:centralresultwithnumbers}).

\begin{figure}
\includegraphics[width = 0.48\textwidth]{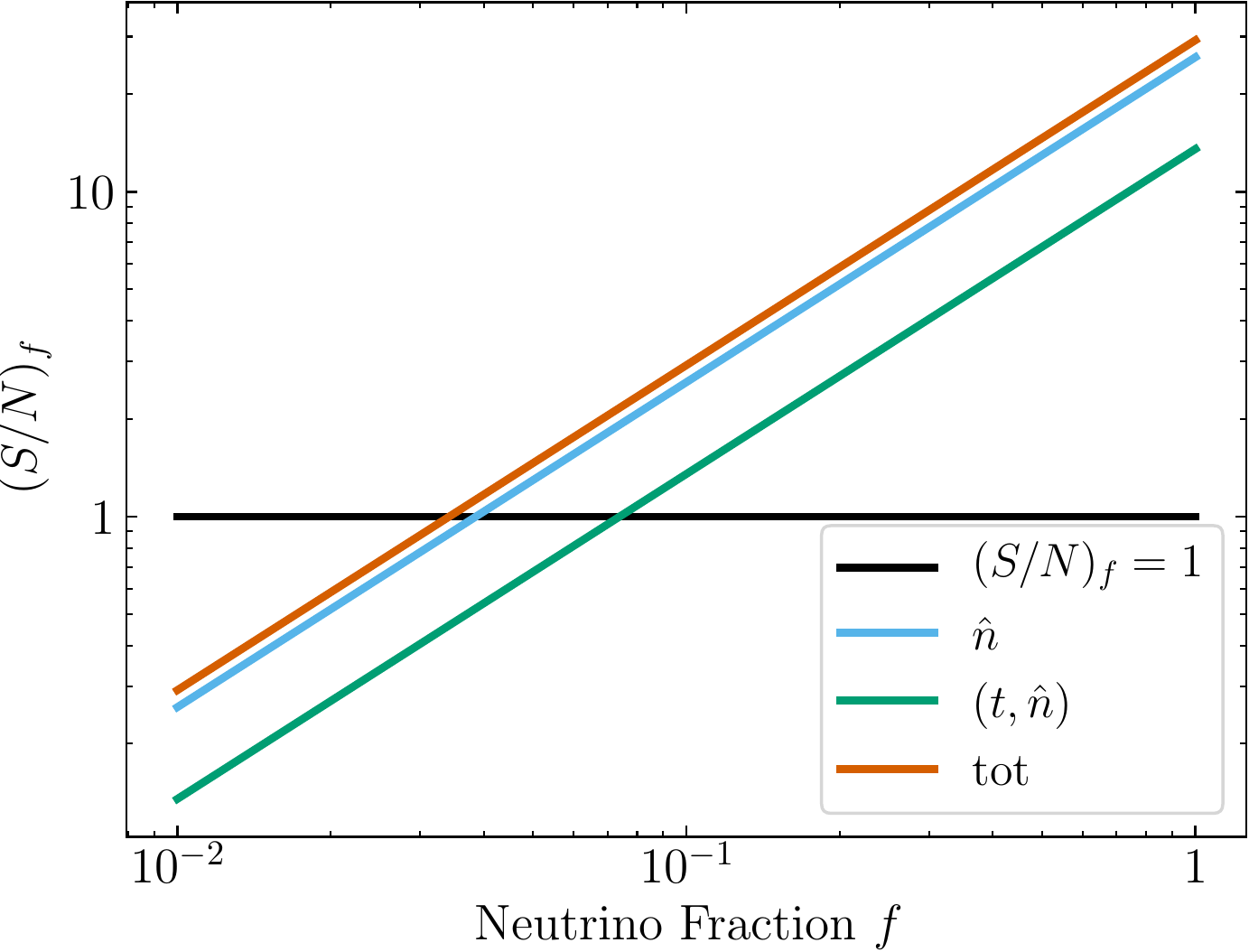}
\caption{The signal-to-noise ratio $(S/N)_f$ for measuring the HEAN neutrino fraction $f$ using both temporal and spatial data from VRO's $i$ band, IceCube, KM3NeT, and Baikal-GVD. Since pure spatial correlation dominates the SNR, the result is not very sensitive to the underlying AGN variability parameters. IceCube contributes $\sim 8\%$ of the total SNR, while KM3NeT and Baikal-GVD each give $\sim 46\%$.}\label{fig:f4cast}
\end{figure}

\subsection{IceCube, KM3NET, and Baikal-GVD}

We now present numerical results including the effective areas
for IceCube, KM3NET, and Baikal-GVD for the regions of sky that
overlap with those surveyed by VRO---to a first approximation,
though, they are all comparable.  We take the angular
resolution of IceCube to be $0.5^\circ$ and those for KM3NET and
Baikal-GVD to be $0.2^\circ$.  The total exposure time is taken
to be 10 years.  We plot the SNR from angular information
alone, from timing, and from the total, in
Fig.~\ref{fig:f4cast}.  The green curve there shows the
signal-to-noise for a measurement where $f$ is inferred only
from a correlation of the neutrino arrival time with AGN
variability, assuming our canonical value for
$\VEV{F^2}/\VEV{F}^2$.

\section{Discussion}\label{sec:disc}

We clarify four assumptions and present two comments. First,
our main assumption is linearity between neutrino number and AGN
bolometric luminosity. Even if linearity holds true, various AGN
may have different proportionality constants due to some
additional specification of AGN class (e.g. this scenario
already occurs with redshift). This will then be encoded in a
change to $\VEV{F^2}/\VEV{F}^2$.  Additional classes in the
variability properties will change the value of
$\VEV{\sigma_{\rm var}^2}$ relative to the value obtained in our
canonical model. It is also possible that the number of neutrinos is
not linear in the AGN's bolometric luminosity, but some power
$\gamma$, with $0\leq \gamma\leq 2$~\cite{1904.06371}. We leave
the investigation of both these cases for future work.

Second, we chose a specific form (the damped random walk) for
the intensity autocorrelation function for AGN.
This form, while applicable to a majority of AGN, has
some exceptions. Changes in the slope, break, as well additional
slopes and breaks, are all required to encapsulate a greater
range of AGN morphologies. However, for our forecast analysis,
such changes will only result in a rescaling of the scaled
variance $\VEV{F^2}/\VEV{F}^2$.  In particular, if the change in the
variability properties shifts the variability timescales outside
of the measurable window allowed by the VRO cadence, then the
prospects to detect a neutrino-AGN temporal correlation will
decrease, while if more power is concentrated in this window,
they may become stronger.

Third, we set the time delay between the neutrino signal and AGN
variability to zero. This was done for simplicity, and in
reality there should be an expected delay depending on where
within the AGN the neutrino was created and where the
variability is sourced. We leave formalizing this description
for future work. 
 
Fourth, we assumed that neutrinos travel along the line of sight unimpeded. The presence of neutrino self-interactions~\cite{2005.05332} can change this description, altering the spatial and temporal coincidence presented here~\cite{1903.05096, 1903.08607}. We also leave exploration of this scenario for future work.

Lastly, measurements of the neutrino fraction $f$ have
covariance with measurements of  AGN variability
parameters. Therefore, in principle the error in measurements of
$f$ should be larger than that presented here. However, given
VRO's precise measurements of an AGN's variability parameters, we
expect such degradation of measurement fidelity to be slight and
our forecast to hold. 

\section{Conclusion}\label{sec:conc}

In this paper, we investigated the prospects to detect an
angular cross-correlation between AGN surveyed by VRO and
energetic neutrinos.  We then discussed further the prospects to
detect a cross-correlation between AGN variability and neutrino
arrival times.

With this aim, we first modelled the spatial cross-correlation
between a single AGN and a population of neutrinos and found a
neutrino-counting measure. More specifically, the contributions
to this correlation were from counting neutrinos sourced by that
AGN and from counting neutrinos with other sources that have
nonzero overlap with that AGN due to angular error. 

AGN may emit electromagnetic radiation along with HEANs, and to
account for this possibility, we also modelled a
temporal-and-spatial cross-correlation. For simplicity, we
assumed that, for each AGN, the number of neutrinos emitted is
proportional to the electromagnetic intensity of that AGN. 

Using both of these correlations, we then forecasted their individual and total abilities to measure the fraction $f$ of HEAN from VRO-observed AGN. The HEANs are detected by a combination of IceCube, KM3NeT, and Bakail-GVD, and we assumed an IceCube-like effective area for each experiment. In accordance with previous work, we took all AGN in the VRO sample to be measured with high signal-to-noise. We thus found that, given 10 years of observation time, temporal and spatial cross-correlations will be able to establish an association between energetic neutrinos and the AGN in VRO even if such AGN contribute only $\sim10\%$ of the neutrino background.  Finally,  given that the background noise scales with $N_{\rm AGN}^{-1/2}$, it should be possible to establish a correlation between neutrinos and some specific subclass of AGN, even if those AGN contribute less than $\sim10\%$ of the neutrino background.

\subsection*{Acknowledgments}

CCS acknowledges the support of the Bill and Melinda Gates
Foundation.  CCS was supported by a National Science Foundation
Graduate Research Fellowship under Grant No.\ DGE-1746891.  This
work was supported in part by the Simons Foundation and by
National Science Foundation grant No.\ 2112699.

\end{document}